\documentclass[aps,prd,twocolumn,superscriptaddress]{revtex4-1}
\usepackage[colorlinks=true, pdfstartview=FitV, linkcolor=red, citecolor=blue, urlcolor=black, pdftitle={},pdfauthor={Tomoya Hayata},pdfsubject={}, pdfkeywords={}]{hyperref}
\usepackage{graphicx}
\usepackage{setspace,bm}
\usepackage{latexsym,amssymb,amsmath,mathrsfs}
\usepackage{color}
\usepackage{times}
\graphicspath{{./figure/}}

\makeatletter
\providecommand \@ifxundefined [1]{%
 \@ifx{#1\undefined}
}%
\providecommand \@ifnum [1]{%
 \ifnum #1\expandafter \@firstoftwo
 \else \expandafter \@secondoftwo
 \fi
}%
\providecommand \@ifx [1]{%
 \ifx #1\expandafter \@firstoftwo
 \else \expandafter \@secondoftwo
 \fi
}%
\providecommand \bibnamefont  [1]{#1}%
\providecommand \bibfnamefont [1]{#1}%
\providecommand \citenamefont [1]{#1}%
\providecommand \href@noop [0]{\@secondoftwo}%
\providecommand \href [0]{\begingroup \@sanitize@url \@href}%
\providecommand \@href[1]{\@@startlink{#1}\@@href}%
\providecommand \@@href[1]{\endgroup#1\@@endlink}%
\providecommand \@sanitize@url [0]{\catcode `\\12\catcode `\$12\catcode
  `\&12\catcode `\#12\catcode `\^12\catcode `\_12\catcode `\%12\relax}%
\providecommand \@@startlink[1]{}%
\providecommand \@@endlink[0]{}%
\providecommand \url  [0]{\begingroup\@sanitize@url \@url }%
\providecommand \@url [1]{\endgroup\@href {#1}{\urlprefix }}%
\providecommand \urlprefix  [0]{URL }%
\providecommand \Eprint [0]{\href }%
\providecommand \doibase [0]{http://dx.doi.org/}%
\providecommand \selectlanguage [0]{\@gobble}%
\providecommand \bibinfo  [0]{\@secondoftwo}%
\providecommand \bibfield  [0]{\@secondoftwo}%
\providecommand \BibitemOpen [0]{}%
\providecommand \EOS [0]{\spacefactor3000\relax}%
\providecommand \BibitemShut  [1]{\csname bibitem#1\endcsname}%
\let\auto@bib@innerbib\@empty

\newcommand{\be}{\begin{equation}}      
\newcommand{\ee}{\end{equation}}      
\newcommand{\bea}{\begin{eqnarray}}      
\newcommand{\eea}{\end{eqnarray}}

\newcommand{\cD}{{\cal D}}

\begin{document}
\title{String confinement in 2-form lattice gauge theory}
\author{Tomoya Hayata}
\affiliation{Nishina Center, RIKEN, Wako 351-0198, Japan}
\author{Arata Yamamoto}
\affiliation{Department of Physics, The University of Tokyo, Tokyo 113-0031, Japan}

\begin{abstract}
We study the confinement between vortex strings in the lattice gauge theory of the dual abelian Higgs model.
The dual lattice gauge theory is described by 2-form gauge field coupled with 1-form gauge field.
We calculate the string-antistring potential from the surface operator of the 2-form gauge field.
The linear confining potential appears in string confinement phase and it disappears in string deconfinement phase.
The phase diagram of the theory is also obtained.
\end{abstract}

\maketitle

\section{Introduction}

A quantum vortex string is a one-dimensional topological soliton.
The existence of vortex strings was experimentally confirmed in superconductors \cite{RevModPhys.82.109} and superfluids \cite{2010arXiv1004.5458T}.
It is also believed to exist in compact stars \cite{Eto:2013hoa} and the Universe \cite{Hindmarsh:1994re}.
The circulation around a vortex string is quantized due to the single-valuedness of a field variable.
The quantized circulation is topologically protected, and thus the vortex string is stable.
The stability ensures the description as quasiparticles, e.g., interaction and dynamics of vortex strings.

The field theory with vortex strings is dual to antisymmetric rank-2 tensor, i.e., 2-form, gauge theory.
The world sheets of vortex strings are described by the surface operator of 2-form gauge field.
This is an analog of the Wilson loop operator in 1-form gauge theory.
The Wilson loop operator corresponds to the world lines of charged particles.
The expectation value of the Wilson loop operator tells us the interaction, e.g., the confinement, between the particles.
Similarly, the interaction between the vortex strings can be investigated from the surface operator of the 2-form gauge field.

The 2-form gauge theory can be non-perturbatively formulated by lattice regularization \cite{Orland:1981ku,Pearson:1981pk,Frohlich:1982gf,Nepomechie:1982rb,Orland:1982fv,Omero:1982hp}.
This is a higher-form generalization of the conventional lattice gauge theory, i.e., the lattice regularization of 1-form gauge theory.
The higher-form lattice gauge theory is sometimes called {\it lattice gerbe theory} \cite{Lipstein:2014vca,Johnston:2014ofa}.
The 2-form lattice gauge theory enables us to study nonperturbative properties of a vortex string from first principles.
Although the vortex string is frequently studied in semi-classical analysis, it misses quantum fluctuation.
First-principle analysis is necessary to take into account quantum fluctuation, e.g., percolation \cite{Baig:1998ui,Kajantie:2000cw,Wenzel:2007uh,MacKenzie:2007ps} and superposition \cite{Hayata:2014kra,Yamamoto:2018vgg}.
Such analysis is particularly important near phase transitions or in finite volumes, where quantum fluctuation is non-negligible.

In this work, we study the confinement phenomenon in 2-form lattice gauge theory.
In Sec.~\ref{sec:LGT}, we review the lattice formulation of 2-form gauge theory.
We consider the 2-form gauge theory coupled with 1-form gauge field.
This theory is dual to the abelian Higgs model in continuum limit~\cite{Quevedo:1996uu}.
The theory exhibits the confinement-deconfinement phase transition of vortex strings.
We confirm that based on  two analyses.
In Sec.~\ref{sec:potential}, we define the Wilson surface operator, and calculate the potential between a string and antistring.
In Sec.~\ref{sec:phase}, we draw the phase diagram of this theory by calculating a susceptibility.

\section{Abelian 2-form lattice gauge theory}
\label{sec:LGT}
We consider novel lattice gauge theory in four-dimensional Euclid spacetime.
The theory contains two kinds of abelian gauge fields: the $1$-form gauge field $A_\mu(x)$ and the 2-form gauge field $B_{\mu\nu}(x)$.
The $1$-form gauge field is defined as a link variable between $x$ and $x+\hat{\mu}$,
\be
U_\mu(x) = e^{iaA_\mu(x)} .
\ee
Here $x$ denotes a site in a hypercubic lattice, $\hat{\mu}$ denotes the unit vector along the $\mu$ direction, and $a$ is lattice spacing.
The 2-form gauge field is defined as a plaquette variable whose vertices are at $x$, $x+\hat{\mu}$, $x+\hat{\nu}$, and $x+\hat{\mu}+\hat{\nu}$,
\be
\Gamma_{\mu\nu}(x) = e^{ia^2B_{\mu\nu}(x)}.
\ee
These variables are U(1) elements.
The lattice action is constructed from these variables as
\begin{equation}
\begin{split}
 S_{\rm lat}&=
 \kappa \sum_{x,\mu<\nu}\;\left(1-\frac{1}{2}\left(\tilde{U}_{\mu\nu}(x)+\tilde{U}^\dagger_{\mu\nu}(x)\right)\right)
 \\
&+ \beta \sum_{x,\mu<\nu<\lambda}\;\left(1-\frac{1}{2}\left(\Gamma_{\mu\nu\lambda}(x)+\Gamma^\dagger_{\mu\nu\lambda}(x)\right)\right),
\end{split}
\label{action:lattice}
\end{equation}
with
\begin{eqnarray}
U_{\mu\nu}(x) &=& U^\dagger_\nu(x)U^\dagger_\mu(x+\hat{\nu})U_\nu(x+\hat{\mu})U_\mu(x) ,
\\
\tilde{U}_{\mu\nu}(x) &=& U_{\mu\nu}(x) \Gamma_{\mu\nu}(x) ,
\\
\Gamma_{\mu\nu\lambda}(x) &=& \Gamma^\dagger_{\lambda\mu}(x+\hat{\nu})\Gamma^\dagger_{\nu\lambda}(x+\hat{\mu})\Gamma^\dagger_{\mu\nu}(x+\hat{\lambda})
\nonumber \\
&&\Gamma_{\lambda\mu}(x)\Gamma_{\nu\lambda}(x)\Gamma_{\mu\nu}(x) .
\end{eqnarray}
Here $\kappa$ and $\beta$ are dimensionless coupling constants.
In the continuum limit, this lattice gauge theory is dual to the abelian Higgs model~\cite{Quevedo:1996uu}.

This theory has two kinds of local gauge symmetry.
The $1$-form gauge transformation is defined with a unitary matrix $e^{i\theta(x)}$ as
\be
U_\mu(x) \rightarrow e^{i\theta(x+\hat{\mu})} U_\mu(x) e^{-i\theta(x)}.
\ee
The minimal gauge-invariant observable is the plaquette operator $U_{\mu\nu}$.
Since the action is written only by $U_{\mu\nu}$, it is manifestly invariant under the $1$-form gauge transformation.
The 2-form gauge transformation is defined with a unitary matrix $e^{i\lambda_\mu(x)}$ as
\be
\Gamma_{\mu\nu}(x) \rightarrow e^{i\lambda_\nu(x)} e^{i\lambda_\mu(x+\hat{\nu})} \Gamma_{\mu\nu}(x) e^{-i\lambda_\nu(x+\hat{\mu})} e^{-i\lambda_\mu(x)}
\ee
and simultaneously
\be
U_\mu(x) \rightarrow e^{i\lambda_\mu(x)} U_\mu(x).
\ee
The minimal gauge-invariant observable is the plaquette operator $\tilde{U}_{\mu\nu}$ and the unit cube operator $\Gamma_{\mu\nu\lambda}$.
Since the action is written by $\tilde{U}_{\mu\nu}$ and $\Gamma_{\mu\nu\lambda}$, it is manifestly invariant under the 2-form gauge transformation.

The expectation value of an operator $\hat{O}$ is given by using the path integral as
\be
\langle\hat{O}\rangle=\frac{\int \cD A_\mu\cD B_{\mu\nu}e^{-S_{\rm lat}}O}{\int \cD A_\mu\cD B_{\mu\nu}e^{-S_{\rm lat}}} .
\ee
Since $e^{-S_{\rm lat}}$ is real and positive, we can compute $\langle\hat{O}\rangle$ on the basis of the standard techniques of Monte Carlo sampling.
In this work, we generated gauge configurations by the heatbath algorithm.
We also adopted the overrelaxation method \cite{PhysRevD.36.515} between the heatbath updates.

\begin{figure}[t]
\begin{center}
\includegraphics[width=.48\textwidth]{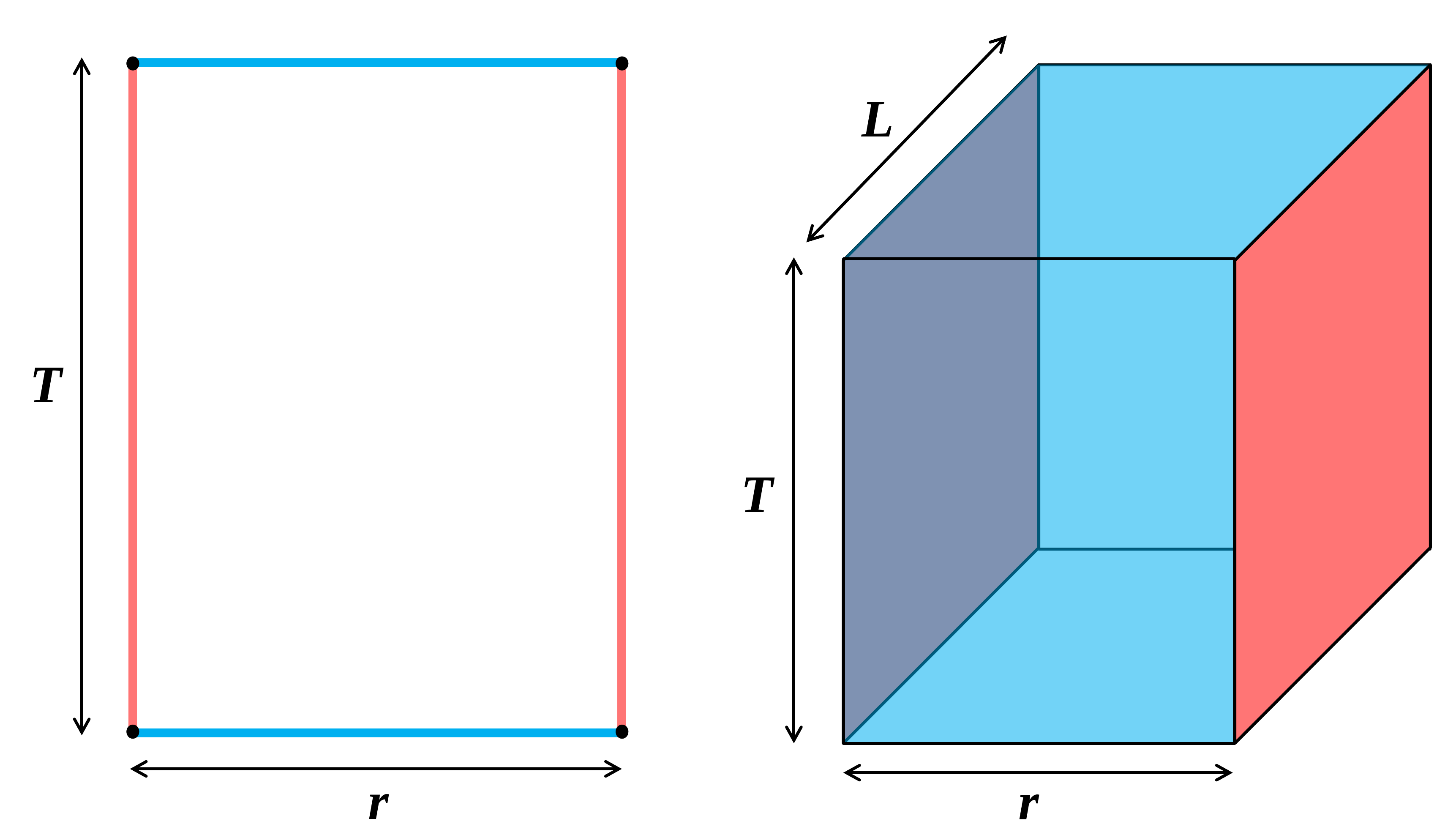}
\caption{
\label{fig1}
Wilson loop (left) and Wilson surface (right).
In the Wilson loop, the two red lines correspond to the trajectories of a particle and an antiparticle.
In the Wilson surface, the two red surfaces correspond to the trajectories of a vortex string and an antivortex string.
The blue lines and surfaces connect these trajectories to make the operators gauge-invariant.
}
\end{center}
\end{figure}

\section{Interstring potential}
\label{sec:potential}

Before discussing the interstring potential in the 2-form lattice gauge theory, let us recall the interparticle potential in $1$-form lattice gauge theory.
Considering a rectangle with the length $r$ and width $T$ in the $\mu$-$\nu$ plane, we can construct a gauge-invariant observable as
\be
\begin{split}
& W_L(r,T) =
\\
&\langle U^\dagger_\nu(x,T)U^\dagger_\mu(x+T\hat{\nu},r)U_\nu(x+r\hat{\mu},T)U_\mu(x,r)\rangle
\end{split}
\ee
with
\be
U_\mu(x,r)=\Pi_{l=0}^{r-1} U_\mu(x+l\hat{\mu}) .
\ee
The schematic figure is shown in Fig.~\ref{fig1}.
This is the so-called Wilson loop.
The Wilson loop corresponds to the world lines of an infinitely-heavy particle and antiparticle.
The ground state energy of the particle-antiparticle pair, i.e., the interparticle potential, is obtained by
\be
aV_q(r)=\lim_{T\to\infty}\log\frac{W_L(r,T)}{W_L(r,T+1)} .
\ee
The area law of the Wilson loop gives the linear confining potential between the particles.
We remark here that this Wilson loop is always zero in our 2-form lattice gauge theory because it is not gauge invariant under the 2-form gauge transformation.
In terms of the abelian Higgs model, this implies the gauge dependence of magnetic monopoles.

This potential calculation can be generalized to the 2-form lattice gauge theory.
Considering a cuboid with the length $r$, width $L$, and height $T$ in the $\mu$-$\nu$-$\lambda$ space, we can construct a gauge-invariant observable as
\be
\begin{split}
& W_S(r,L,T) =
\\
&\langle \Gamma^\dagger_{\lambda\mu}(x+L\hat{\nu},T,r)\Gamma^\dagger_{\nu\lambda}(x+r\hat{\mu},L,T)\Gamma^\dagger_{\mu\nu}(x+T\hat{\lambda},r,L)
\\
&\Gamma_{\lambda\mu}(x,T,r)\Gamma_{\nu\lambda}(x,L,T)\Gamma_{\mu\nu}(x,r,L)\rangle
\end{split}
\ee
with
\be 
\Gamma_{\mu\nu}(x,r,L)= \Pi_{l=0}^{r-1} \Pi_{m=0}^{L-1} \Gamma_{\mu\nu}(x+l\hat{\mu}+m\hat{\nu}),
\ee
as shown in Fig.~\ref{fig1}.
In the limit of $L \to \infty$, this Wilson surface corresponds to the world sheets of an infinitely-heavy and infinitely-long string and antistring.
The ground state energy is obtained by
\be
aV_S(r)=\lim_{T,L\to\infty}\log\frac{W_S(r,L,T+1)W_S(r,L+1,T)}{W_S(r,L,T)W_S(r,L+1,T+1)} .
\ee
This is the interstring potential per unit length.
The volume law of the Wilson surface gives the linear confining potential between the strings.

\begin{figure}[t]
\centering
\includegraphics[width=.48\textwidth]{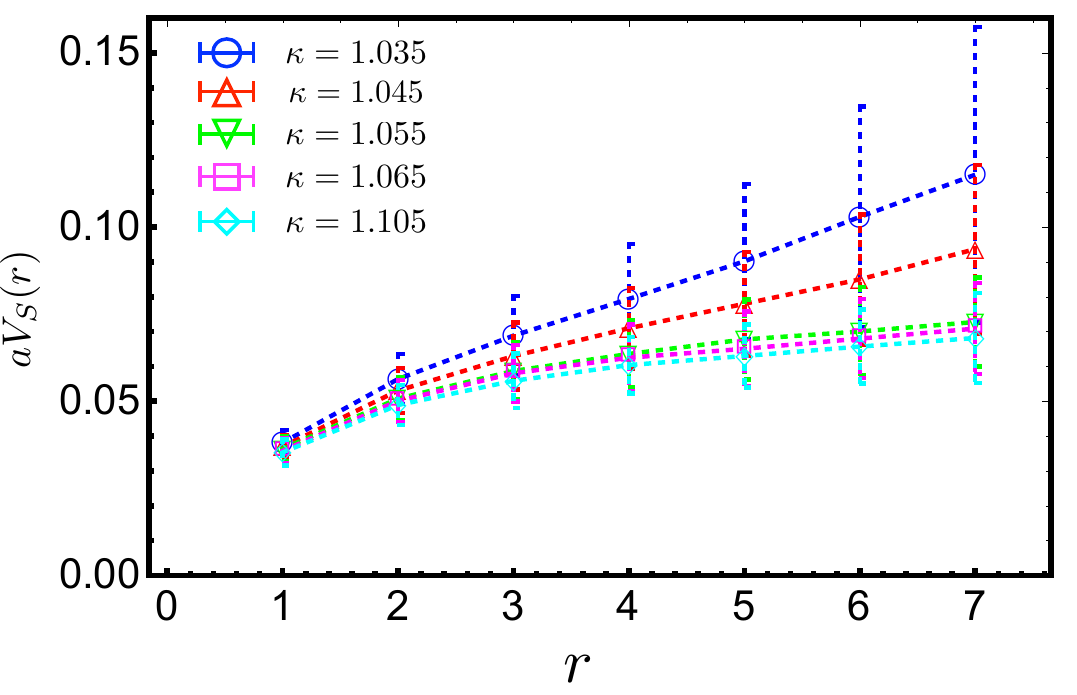}
\caption{\label{fig2}
Interstring potential with $\beta=4.1$, and various $\kappa$.
The confining linear potential is clearly seen in the confinement phase at small $\kappa$, while it disappears in the deconfinement phase at large $\kappa$.
}
\end{figure}

We computed the Wilson surface in the Monte Carlo simulation.
The lattice volume is $V=16^4$.
The APE smearing was employed to compute the interstring potential efficiently~\cite{ALBANESE1987163}.
The result is shown in Fig.~\ref{fig2}.
At small $\kappa$, we clearly see the linearly rising potential.
This is interpreted as the confinement between a string and antistring.
We call it ``string confinement'' to distinguish it from the ordinary confinement between point particles.
The linear potential disappears as $\kappa$ increases, which means ``string deconfinement''.
This result suggests a new type of the phase transition characterized by the (de-)confinement of extended objects.
This is the main result of this paper.

\begin{figure}[t]
\centering
\includegraphics[width=.48\textwidth]{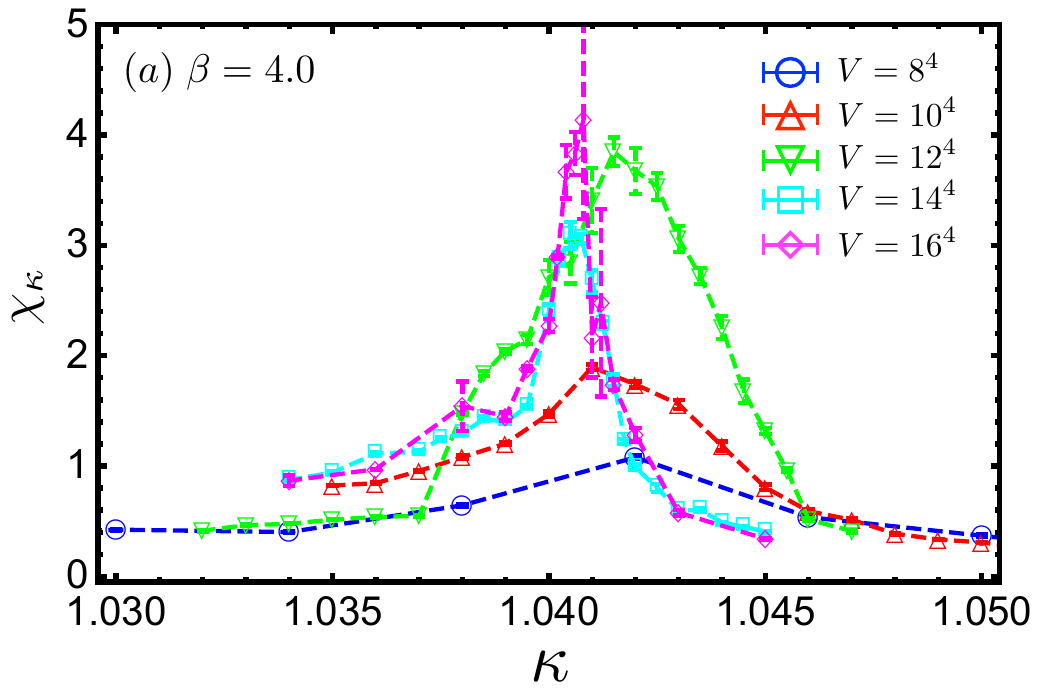}
\centering
\includegraphics[width=.48\textwidth]{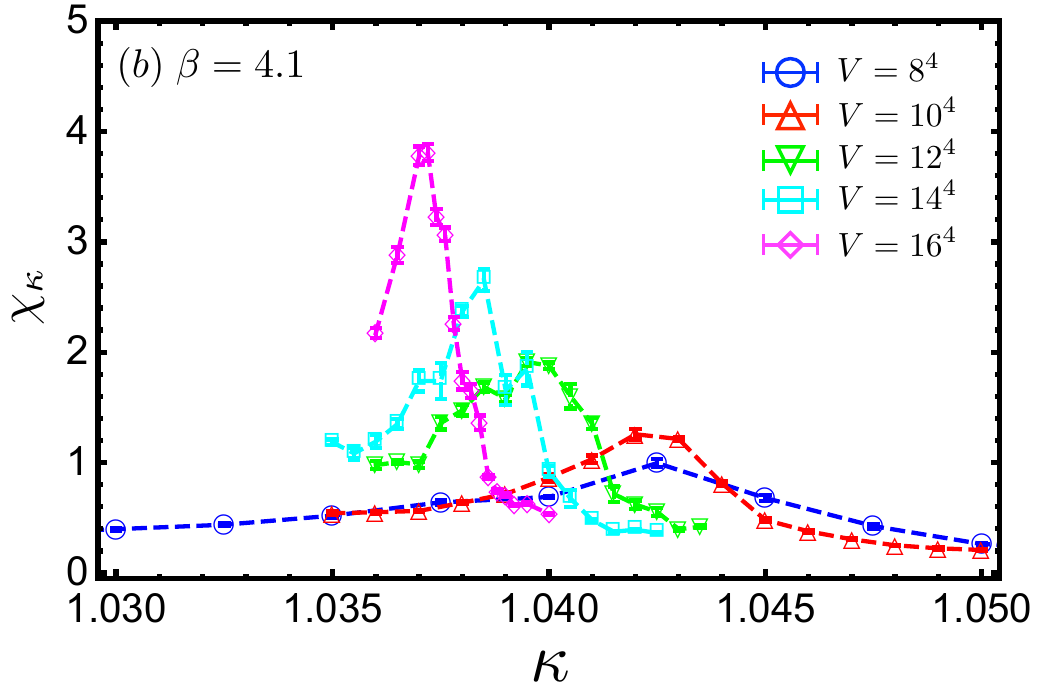}
\caption{\label{fig3}
Volume-dependence of the susceptibility $\chi_\kappa$: ($a$) $\beta=4.0$ below the critical point and ($b$) $\beta=4.1$ above the critical point.
The last three volumes are used for the finite-size scaling analysis to determine the critical $\kappa$ in the infinite volume limit. 
}
\end{figure}

The numerical calculation was done in finite interstring distance.
Does the linear confining potential persist in infinite distance?
To answer this question, let us introduce another gauge-invariant observable
\be
W'_S(r,L,T) = \langle \tilde{U}_{\nu\lambda}(x,L,T) \tilde{U}_{\nu\lambda}^\dagger(x+r\hat{\mu},L,T) \rangle
\ee
with
\be
\tilde{U}_{\mu\nu}(x,r,L) = \Pi_{l=0}^{r-1} \Pi_{m=0}^{L-1} \tilde{U}_{\mu\nu}(x+l\hat{\mu}+m\hat{\nu}) .
\ee
The physical picture of this observable is the world sheets of a string and antistring attached with the 1-form gauge field.
Since these string and antistring are independently gauge invariant, they are not confined but weakly coupled.
In this theory, $W_S(r,L,T)$ and $W_S'(r,L,T)$ have the same quantum number.
The two states, the confined and non-confined states, are mixed.
Since the ground state is the one with lower energy, the non-confined state will be favored in large distance.
Therefore the corresponding potential will not be linear but constant.
This is the same as the string breaking in quantum chromodynamics.
The potential between a quark and antiquark is linear in short distance but constant in long distance.
The confining string connecting the quark and antiquark is broken by dynamical quark-antiquark pair creation.
In our case, the confining surface connecting a string and antistring will be broken by the 1-form gauge field in long range limit.
This should be called ``surface breaking''.
The critical distance $r_c$ where the surface breaking takes place can be estimated by the energy balance relation $V_S(r_c) = 2M$.
Here $M$ is the mass of one gauge-invariant string defined by $ \lim_{T,L\to\infty} \langle \tilde{U}_{\nu\lambda}(x,L,T) \rangle \propto \exp(-MLT)$.
In this simulation, typical values are $2aM \simeq 0.38$ at $\beta =4.1$ and $\kappa =1.035$, $2aM \simeq 0.14$ at $\beta =4.1$ and $\kappa =1.065$, and $2aM \simeq 0.13$ at $\beta =4.1$ and $\kappa =1.105$.
We see that the data in Fig.~\ref{fig2} is below the critical distance.
The direct simulation of the surface breaking would be an interesting future work.
In principle, both $W_S(r,L,T)$ and $W_S'(r,L,T)$ can give the same correct result in the limit $T,L\to\infty$.
In practice, however, special treatment is necessary for technical reasons \cite{Bali:2005fu}.

\section{Phase diagram}
\label{sec:phase}

We draw the phase diagram of this theory in the parameter space of $\kappa$ and $\beta$.
We calculated the susceptibility
\begin{equation}
  \chi_\kappa= \frac{1}{V} \left\langle \left( \frac{\partial S_{\rm lat}}{\partial \kappa}-\left\langle\frac{\partial S_{\rm lat}}{\partial \kappa}\right\rangle \right)^2 \right\rangle
\end{equation}
to determine the position and the order of the phase transition.
The parallel tempering was employed to compute $\chi_\kappa$~\cite{doi:10.1143/JPSJ.65.1604}.
As examples,  we show the volume-dependence of $\chi_\kappa$ at $\beta=4.0$ and $4.1$ in Fig~\ref{fig3}.
At $\beta=4.1$, we observed the double peak structure implying metastable states and the volume growth of the susceptibility $\chi_{\kappa}\propto V$.
Both are strong evidences of the first-order phase transition. 

\begin{figure}[t]
\centering
\includegraphics[width=.48\textwidth]{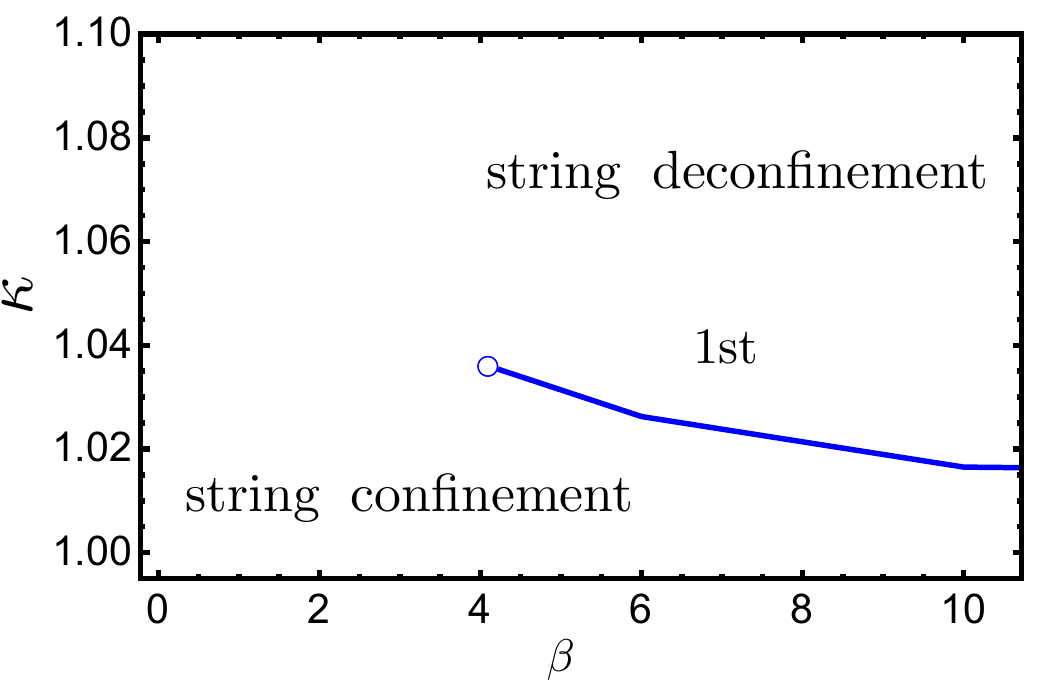}
\caption{\label{fig4}
Phase diagram of the 2-form lattice gauge theory.
The blue curve is the first-order phase-transition line.
}
\end{figure}

The obtained phase diagram is shown in Fig. \ref{fig4}.
The finite-size scaling analysis was done for three lattice volumes $V=12^4$, $14^4$, and $16^4$.
There are two phases: the string confinement phase and the string deconfinement phase.
In small $\beta$, the two phases are smoothly connected by a crossover.
In large $\beta$, the two phases are separated by a first-order phase transition.
In the limit of $\beta \to \infty$, the lattice action \eqref{action:lattice} reduces to the conventional 1-form compact U(1) gauge action.
There must be a first-order phase transition in this limit.
This is consistent with our observation.
The first-order phase-transition line ends at a critical point.
The position of the critical point was estimated as $\kappa_c \simeq 1.036$ and $\beta_c \simeq 4.1$.

This is the phase diagram in the case of the unit charge.
The theory can be generalized to the charge $N$ representation by replacing $\tilde{U}_{\mu\nu} = U_{\mu\nu} \Gamma_{\mu\nu} \to U_{\mu\nu} (\Gamma_{\mu\nu})^N$ in the lattice action \eqref{action:lattice}.
This is analogous to the phase diagram with the charge-$N$ Higgs field \cite{Fradkin:1978dv}.
The phase diagram for the charge $N>1$ would be more interesting because the theory has $Z_N$ topological order.

\begin{acknowledgements}
The authors thank Tin~Sulejmanpasic for pointing out the possibility of surface breaking.
T.~H. thanks Takumi~Doi, Sinya~Gongyo, and Yuta~Kikuchi for helpful comments.
T.~H. was supported by RIKEN special postdoctoral program.
A.~Y.~was supported by JSPS KAKENHI Grant Number 19K03841. 
\end{acknowledgements}

\bibliography{paper}

\end{document}